\documentclass[12pt]{article}
\usepackage{enumerate,amssymb}
\renewcommand{\a}{\alpha}
\renewcommand{\b}{\beta}
\newcommand{\g}{\gamma}           
\renewcommand{\d}{\delta}

\newcommand{\la}{\lambda}

\newcommand{\s}{\sigma}           
         
\newcommand{\f}{{\phi}}

\newcommand{\be}{\begin{equation}}
\newcommand{\ee}{\end{equation}}
\newcommand{\eqn}[1]{\label{#1}\end{equation}}

\newcommand{\bea}{\begin{eqnarray}}
\newcommand{\eea}{\end{eqnarray}}
\newcommand{\eqan}[1]{\label{#1}\end{eqnarray}}

\newcommand{\ba}{\begin{array}}
\newcommand{\ea}{\end{array}}

\newcommand{\nn}{\nonumber}

\begin{document}

\begin{center}
{\bf   String Corrected Supergravity; A Complete and Consistent Non-Minimal Solution}\\[14mm]

D. O'Reilly\\

{\it Physics Department, The Graduate School and University
Center\\365 Fifth Avenue
New York, NY 10016-4309\\ mailto: doreilly@gc.cuny.edu and dunboyne@vzavenue.net}\\[6mm]

\end{center}
\vbox{\vspace{3mm}}

\begin{abstract}  We complete the solution to string corrected (deformed),
 D=10, N=1 Supergravity as the non-minimal low energy limit of string theory.
 We reaffirm a previously given solution, and we make important corrections to that solution.
 We solve what was an apparently
 intractable Bianchi identity in superspace, and we introduce a new
 important modification to the known first order results. In so
 doing we show that this approach to string corrected supergravity
 is indeed a consistent approach and we pave the way for many
 applications of the results.

 \vbox{\vspace{1mm}}

\end{abstract}

\newpage

\section{Introduction}

The route to finding a manifestly supersymetric theory of D= 10, N=1
supergravity at second order in the string slope parameter has
encountered many difficulties over the years. Some years ago a
solution to D=10, N=1 Supergravity as the low energy limit of String
Theory was given at first order in the string slope parameter,
\cite{1}. It was recently re-calculated \cite{1}/(2004). In a sense
this was a minimal solution. This approach was founded on what we
now choose to call the scenario of Gates and collaborators, (see
\cite{1}, \cite{2}, and references therein). Other varied approaches
are nowadays pursued, however the power of this older approach is
currently being vindicated,\cite{1}. A partial second order solution
was recently given in \cite{3} and \cite{4}. It was incomplete and
therefore in doubt due to an unsatisfactory assumption in the
curvature sector, as well as a calculational error. Here we reaffirm
that that solution is correct up to a curvature. We then show that
the results obtained satisfy the problem curvature, equation (3). We
achieve this through introducing a new and important condition on
$R^{(1)}{}_{ab\a}{}^{\g}$, a quantity previously undefined. This
result also modifies the old first order case. The difficulties that
prevented completely closing the Bianchi identities at second order
are fully overcome. We complete the set of equations that
consistently satisfy all Bianchi identities. As the work in itself
is lengthy we leave finding the equations of motion and other
applications for another paper. We do not list results which are
explicitly solved by Bianchi identities such as $H^{(2)}{}_{abc}$.

For this approach it is required that we solve the Bianchi
identities for D=10 N=1 Supergravity in Superspace at second order
in the slope parameter, in the presence of the Lorentz Chern Simmons
Form, and the so called Beta Function Favored Constraints, $(\b
FF$). This approach has been detailed to first order in \cite{1},
and to second order in \cite{3} and \cite{4}, so we will not recount
it here. We show that all results fall neatly into place in a very
elegant way, therefore further vindicating the whole original
scenario. We note here that it appears also to work consistently at
third order as we have proceeded to that order and that is left for
another letter.

\section{Review of Solution and Notation}

The Bianchi identities in Superspace are as follows

\bea[[\nabla_{[A},\nabla_{B}\},\nabla_{C)}\}=0\eea \

Here we have switched off Yang Mills fields and the commutator is
given by

 \bea
[\nabla_{A},\nabla_{B}\}=T_{AB}{}^{C}+\frac{1}{2}R_{ABd}{}^{e}M_{e}{}^{d}
\eea

This generates many identities, and a solution must be found in such
a way that all of them are satisfied simultaneously. A small
alteration in one solution will change the whole picture. Most of
the resulting identities are listed in \cite{1} and \cite{4}, so we
will not list them here. The second order solution given in parts in
\cite{3} and \cite{4} was, to some extent based on an anzatz for the
so called X tensor as well as extensive algebraic manipulations. The
necessity for introducing the X tensor was predicted by Gates et.
al., \cite{1}. In \cite{3}, and \cite{4}, the following Bianchi
identity was not properly solved.

 \bea
T_{(\a\b|}{}^{\la}R_{|\g)\la}{}_{de}~-~T_{(\a\b|}{}^{g}R_{|\g)}{}_{gde}~
-~\nabla_{(\a|}R_{\b\g)}{}_{de}~=~0\eea

It is crucial to show that the torsions and curvature already found
(7), (11) and (20),  satisfy this identity otherwise the whole set
of equations is in doubt. Also $R^{(2)}{}_{\g gde}$ is required to
complete the set.
 Various ideas such as finding a new X tensor, imposing constraints on the spinor
derivative $\nabla_{\a}\chi$ at second order or adjusting the super
current $A_{abc}$ were previously fruitlessly considered.

In this paper we find a consistent solution. We also point out that
equation (58) in reference \cite{3} (or equation (115) in reference
\cite{4}) is wrong.

In order to avoid a proliferation of terms we maintain the same
notation and conventions as in \cite{1} and \cite{4}, but to avoid
relisting the first order results we denote all quantities by order
in the slope parameter as follows

\bea
R_{ABde}=R^{(0)}{}_{ABde}+R^{(1)}{}_{ABde}+R^{(2)}{}_{ABde}+...\nn\eea
\newline
\bea
T_{AD}{}^{G}=T^{(0)}{}_{AD}{}^{G}+T^{(1)}{}_{AD}{}^{G}+T^{(2)}{}_{AD}{}^{G}...\nn\eea
The numerical superscript refers to the order of the quantity. In
this work we make some improvements to the notation in references
\cite{3} and \cite{4}. For convenience we also have the following
quantity

\bea \Omega^{(1)}{}_{gef}=L^{(1)}{}_{gef}-\frac{1}{4}A^{(1)}{}_{gef}
\eea and its spinor derivative

\bea \Omega^{(1)}{}_{\a
gef}=\nabla_{\g}\{\L^{(1)}{}_{gef}-\frac{1}{4}A^{(1)}{}_{gef}\} \eea

A crucial input at first order is that for the super-current
$A^{(1)}{}_{gef}$.  The choice made for on-shell conditions in
\cite{1} and hence also \cite{3} and \cite{4}, is as follows

\bea A^{(1)}{}_{gef}=+i\g \s_{gef \epsilon \tau}T^{mn
\epsilon}T_{mn}{}^{\tau} \eea

In \cite{3} and \cite{4},  we proposed the form of the X tensor to
be as follows \bea T^{(2)}{}_{\a\b}{}^{d}~=\s^{pqref}{}_{\a\b}
X_{pqrefd}=-\frac{i\g}{6}\s^{pqref}{}_{\a\b}
H^{(0)}{}^{d}{}_{ef}A^{(1)}{}_{pqr}\eea

Coupled with this we also have a conventional constraint which may
or may not be imposed to all orders. We have

 \bea T_{\a b}{}^{\d}=-\frac{1}{48}\s_{b\a \la}\s^{pqr
\la\d}A_{pqr}\eea

If we impose this at second order we have a result that relates this
torsion at second order to the super current.

 \bea T^{(2)}{}_{\a b}{}^{\d}=-\frac{1}{48}\s_{b\a \la}\s^{pqr
\la\d}A^{(2)}{}_{pqr}\eea

However we may relax this constraint also. We will consider this
option in reconsidering the solution to equation (13).

A fundamental result which was used in every Bianchi identity and
which is very lengthy to derive is the following

\bea
~T^{(0)}{}_{(\a\b|}{}^{\la}\s^{pqref}{}_{|\g)\la}A^{(1)}{}_{pqr}H^{(0)}{}_{def}-
\s^{pqref}{}_{(\a\b|}H^{(0)}{}_{def}\nabla_{|\g)}A^{(1)}{}_{pqr}\nn\\
=-24 \s^{g}{}_{(\a\b|}H^{(0)}{}_{d}{}^{ef}[\Omega^{(1)}{}_{|\g)
gef}]\eea

We note however in this paper that this result can be arrived at
indirectly by using the first order results found in \cite{1}, in
conjunction with the Bianchi identity (3).
 \section{Torsions Solutions}

We found from the H sector Bianchi identities that the following
dimension one half torsion is given uniquely by

\bea
T^{(2)}{}_{\a\b}{}^{\la}~=~-~\frac{i\g}{12}\s^{pqref}{}_{\a\b}A^{(1)}{}_{pqr}T_{ef}{}^{\la}
\eea It was then shown that together with the proposed X tensor
anzatz (7), as well as equation (8) and other observations and
results, that the H sector Bianchi identities as listed in \cite{1},
\cite{2} could be solved. Also solved was the torsions (10), below.

\bea
T_{(\a\b|}{}^{\la}T_{|\gamma)\la}{}^{d}~-~T_{(\a\b|}{}^{g}T_{|\gamma)g}{}^{d}~
-~\nabla_{(\a|}T_{\b\g)}{}^{d}~=~0 \eea

These results also offer a solution to the following

\bea
T_{(\a\b|}{}^{\la}T_{|\g)\la}{}^{\d}~-~T_{(\a\b|}{}^{g}T_{|\g)g}{}^{\d}
~-~\nabla_{(\a|}T_{|\b\g)}{}^{\d} -~\frac{1}{4}R_{(\a\b|
de}\s^{de}{}_{|\g)}{}^{\d} ~=~0 \eea

However consideration must be given here as to whether or not to
impose the constraint (9). Either way we can solve the identity. It
is important to note that imposing this constraint results in a null
term in (13),

\bea
T^{(0)}{}_{(\a\b|}{}^{g}T^{(2)}{}_{|\g)g}{}^{\d}=i\s_{(\a\b|}{}^{g}\{-\frac{1}{48}\s_{g|\g)
\la}\s^{pqr \la\d}A^{(2)}{}_{pqr}\} \eea

This is due to the fact that \bea
\s_{(\a\b|}{}^{g}\s_{g|\g)\la}=0\eea

We find the second order solutions to (12) to be given by (7) and
the following

\bea \s^{g}{}_{(\a\b|}T^{(2)}{}_{|\g)gd}
=4\g\s^{g}{}_{(\a\b|}\Omega_{|\g) gef}H^{(0)}{}_{d}{}^{ef}-
\frac{i\g}{6}\s^{g}{}_{(\a\b|}\s^{pqre}{}_{g|\g)\f}A^{(1)}{}_{pqr}T^{(0)}{}_{de}{}^{\f}
\eea

The lengthy extracted and symmetrized equation is listed in
\cite{3}, and \cite{4}. In equation (13), we notice the occurrence
of the term

\bea -\nabla_{(\a|}T^{(0)}{}_{|\b\g)}{}^{\d}{}^{~[Order (2)]}=
[2\d_{(\a|}{}^{\d}\d_{|\b)}{}^{\la}+\s^{g}{}_{(\a\b|}\s_{g}{}^{\d
\la}]\nabla_{|\g)}\chi_{\la}{}^{(2)}\eea

If we impose the constraint (14), then this term must be retained as
being non zero. If we relax the constraint (14), then we may include
this as an extra constraint as follows, \bea
[2\d_{(\a|}{}^{\d}\d_{|\b)}{}^{\la}+\s^{g}{}_{(\a\b|}\s_{g}{}^{\d
\la}]\nabla_{|\g)}\chi_{\la}{}^{(2)}=0\eea

Here we chose to relax the constraint. Hence in so doing we find for
the solution of (13) after some algebra and neat cancelations,

\bea T^{(2)}{}_{\g g}{}^{\d} =2\g~T^{(0)}{}^{ef}{}^{\d}
\Omega^{(1)}{}_{\g gef} \eea

And \bea R^{(2)}{}_{ \a\b de}
 =-~\frac{i\g}{12}\s^{pqref}{}_{\a\b}A^{(1)}{}_{pqr}R^{(0)}{}_{ef de}\eea

We now must show that all of the above found results satisfy (3).

\section{New Solution for $R^{(2)}{}_{\la gde}$ }
We must show that we can close equation (3) using the results (7),
(11), and (20). As mentioned, in references \cite{3} and \cite{4}
the curvature (3) was not properly solved. In fact there existed
terms which seemed at first to predict serious problems for the
entire scenario. The mentioned various approaches did not work, nor
was there any way to manipulate the terms using the sigma matrix
algebra. Eventually the following procedure provided a confident and
elegant solution. At second order the Bianchi identity (3) becomes

\bea T^{(0)}{}_{(\a\b|}{}^{\la}R^{(2)}{}_{|\g)\la}{}_{de}+
T^{(2)}{}_{(\a\b|}{}^{\la}R^{(0)}{}_{|\g)\la}{}_{de}~
-T^{(0)}{}_{(\a\b|}{}^{g}R^{(2)}{}_{|\g)g}{}_{de}~-
~T^{(2)}{}_{(\a\b|}{}^{g}R^{(0)}{}_{|\g)g}{}_{de}\nn\\
-~\nabla_{(\a|}[R^{(0)}{}_{|\b\g)}{}_{de}^{Order
(2)}+~R^{(1)}{}_{|\b\g)}{}_{de}^{Order
(2)}+~R^{(2)}{}_{|\b\g)}{}_{de}^{Order (2)}]~=~0\nn\\\eea

 Using the results we found, (7), (11) and (20), we arrive at

 \bea -i\s^{g}{}_{(\a\b|}R^{(2)}{}_{|\g)}{}_{gde}
 +T^{(0)}{}_{(\a\b|}{}^{\la}[
 -~\frac{i\g}{12}\s^{pqrab}{}_{|\g)\la}A^{(1)}{}_{pqr}R^{(0)}{}_{abde}]\nn\\
 -~\frac{i\g}{12}\s^{pqrab}{}_{(\a\b|}A^{(1)}{}_{pqr}T_{ab}{}^{\la}R^{(0)}{}_{|\g)\la}{}_{de}
~+~\frac{i\g}{6}\s^{pqrab}{}_{(\a\b|}
H^{(0)}{}^{g}{}_{ab}A^{(1)}{}_{pqr}R^{(0)}{}_{|\g)}{}_{gde}\nn\\
-\nabla_{(\g|}\{-2i\s^{g}{}_{|\a\b)}\Pi^{(0)+(1)}{}_{gde}+
\frac{i\g}{24}\s^{pqr}{}_{de}{}_{|\a\b)}A^{(1)}{}_{pqr}\nn\\
-~\frac{i\g}{12}\s^{pqrab}{}_{|\a\b)}A^{(1)}{}_{pqr}R^{(0)}{}_{abde}\}=0
\eea

Here we encounter second order contributions from zeroth order parts
but in solvable form. We define

 \bea
 \Pi_{g}{}^{ef}=L{}_{g}{}^{ef}
-~\frac{1}{8}A_{g}{}^{ef} \eea

Now again using our key relation, (10)  we obtain

\bea -i\s^{g}{}_{(\a\b|}R^{(2)}{}_{|\g)}{}_{gde}+2i\g
\s^{g}{}_{(\a\b|}R^{(0)}{}_{abde}[\Omega^{(1)}{}_{|\g) gab}]
-\nabla_{(\g|}\{-2i\s^{g}{}_{|\a\b)}\Pi^{(0)+(1)}{}_{gde}\}\nn\\
-~\frac{i\g}{12}\s^{pqrab}{}_{(\a\b|}A^{(1)}{}_{pqr}T_{ab}{}^{\la}R^{(0)}{}_{|\g)\la}{}_{de}
+~\frac{i\g}{6}\s^{pqrab}{}_{(\a\b|}H^{(0)}{}^{g}{}_{ab}A^{(1)}{}_{pqr}R^{(0)}{}_{|\g)}{}_{gde}\nn\\
+~\frac{i\g}{12}\s^{pqrab}{}_{(\a\b|}A^{(1)}{}_{pqr}[\nabla_{|\g)}R^{(0)}{}_{abde}]~
-~\frac{i}{24}\s^{pqr}{}_{de}{}_{(\a\b|}[\nabla_{|\g)}A^{(1)}{}_{pqr}^{Order
(2)}]\nn\\
\eea

Of particular concern and interest is the last term in (22). One
possible approach to eliminating this term is that taken in \cite{3}
and \cite{4}. However here we now disagree with that approach. Hence
the problem terms will still remain. It was thought that a possible
modification of $A^{(1)}{}_{pqr}$, or a contribution from
$A^{(2)}{}_{pqr}$ would be necessary. These approaches are now seen
also to be unnecessary.

In advance we anticipate that the solution will be as follows

\bea +i\s^{g}{}_{(\a\b|}R^{(2)}{}_{|\g)}{}_{gde}= 2i\g
\s^{g}{}_{(\a\b|}R^{(0)}{}_{abde}[\Omega^{(1)}{}_{|\g) g}{}^{ab}]
+\nabla_{(\g|}\{2i\s^{g}{}_{|\a\b)}\Pi^{(0)+(1)}{}_{gde}\}{}^{~Order (2)}\nn\\
\eea And

 \bea
-~\frac{i\g}{12}\s^{pqrab}{}_{(\a\b|}A^{(1)}{}_{pqr}T_{ab}{}^{\la}R^{(0)}{}_{|\g)\la}{}_{de}
+~\frac{i\g}{6}\s^{pqrab}{}_{(\a\b|}H^{(0)}{}^{g}{}_{ab}A^{(1)}{}_{pqr}R^{(0)}{}_{|\g)}{}_{gde}\nn\\
+~\frac{i\g}{12}\s^{pqrab}{}_{(\a\b|}A^{(1)}{}_{pqr}[\nabla_{|\g)}R^{(0)}{}_{abde}]~
-~\frac{i}{24}\s^{pqr}{}_{de}{}_{(\a\b|}[\nabla_{|\g)}A^{(1)}{}_{pqr}^{Order
(2)}]=0 \eea

We need to show that (26) does in fact vanish. We must begin with
the Bianchi identity that gives the spinor derivative of
$T_{kl}{}^{\tau}$.

\bea
\nabla_{\g}T_{kl}{}^{\tau}=T_{\g[k|}{}^{\la}T_{\la|l]}{}^{\tau}+
T_{\g[k}{}^{g}T_{g|l]}{}^{\tau}+T_{kl}{}^{\la}T_{\la
\g}{}^{\tau}+T_{kl}{}^{g}T_{g\g}{}^{\tau}-\nabla_{[k|}T_{|l]\g}{}^{\tau}
-R_{kl\g}{}^{\tau}\eea
 At first order this simplifies to

 \bea \nabla_{\g}T_{kl}{}^{\tau}{}^{Order (1)}=-R^{(1)}{}_{kl\g}{}^{\tau}+
 T^{(1)}{}_{kl}{}^{\la}T^{(0)}{}_{\la
\g}{}^{\tau}\nn\\+\frac{1}{48} [2H^{(0)}{}_{kl}{}_{g}\s^{g}{}_{\g
\la}\s^{pqr\la\tau}A^{(1)}{}_{pqr}
-\s_{[k|\g\la}\s^{pqr\la\tau}{}(\nabla_{|l]}A^{(1)}{}_{pqr})]\eea

We now write the last term in (26), using the ten dimensional metric
so that the unsolved part becomes

 \bea
-\frac{i}{12}\s^{pqrab}{}_{(\a\b|}\{\g
A^{(1)}{}_{pqr}[T_{ab}{}^{\la}R^{(0)}{}_{|\g)\la}{}_{de}
+T^{(0)}_{ab}{}^{g}R^{(0)}{}_{|\g)}{}_{gde}-\nabla_{|\g)}R^{(0)}{}_{abde}]~\nn\\
+\frac{1}{2}\eta_{ad}~\eta_{be}\nabla_{|\g)}A^{(1)}{}_{pqr}^{~(Order
(2))}\}=0 \eea

Using the definition of $A^{(1)}{}_{pqr}$, (6), gives therefore
 \bea
+\frac{\g}{12}\s^{pqrab}{}_{(\a\b|}\s_
{pqr\epsilon\tau}T^{kl\epsilon} \{\g
T_{kl}{}^{\tau}[T_{ab}{}^{\la}R^{(0)}{}_{|\g)\la}{}_{de}
+T^{(0)}_{ab}{}^{g}R^{(0)}{}_{|\g)}{}_{gde}-\nabla_{|\g)}R^{(0)}{}_{abde}]~\nn\\
+\eta_{ad}\eta_{be}\nabla_{|\g)}T_{kl}{}^{\tau}\}=0 \eea

We now use equation (25) and the properties of the sigma matrices.
After some algebra we obtain an extremely interesting condition on
$R^{(1)}{}_{kl\g}{}^{\tau}$. We find

\bea
R^{(1)}{}_{kl\g}{}^{\tau}=\{+\frac{\g}{100}T{}_{kl}{}^{\tau}[T_{mn}{}^{\la}R^{(0)}{}_{\g\la}{}^{mn}
+T^{(0)}{}_{mn}{}^{g}R^{(0)}{}_{|\g)}{}_{g}{}^{mn}-\nabla_{ \g
}R^{(0)}{}_{mn}{}^{mn}]\nn\\+
 T^{(1)}{}_{kl}{}^{\la}T^{(0)}{}_{\la
\g}{}^{\tau}
+4i\g[T_{mn}{}^{\la}T^{mn\tau}H^{(0)}{}_{klg}\s^{g}{}_{\g
\la}-\s_{[k|\g \la}T_{mn}{}^{\la}\nabla_{|l]}T^{mn\tau}] \} \eea

This can now be added to the list of first order results quoted in
\cite{1}. It assumes a correction $T^{(1)}{}_{kl}{}^{\la}$ which may
itself be complicated.
 $R^{(1)}{}_{kl\g}{}^{\tau}$ was not defined in \cite{1}. Furthermore
the curvature (3) is neatly solved. We obtain

\bea R^{(2)}{}_{\g}{}_{gde}= 2\g R^{(0)}{}_{abde}[\Omega^{(1)}{}_{\g
g}{}^{ab}]
+\nabla_{\g}\{\Pi^{(0)+(1)}{}_{gde}\}{}^{Order (2)}\nn\\
\eea

 The following Bianchi identity  also includes $R^{(2)}{}_{\a bde}$.

\bea
 \frac{1}{4}R_{(\a|}{}_{amn}\s^{mn}{}_{|\b)}{}^{\g}
+T_{\a\b}{}^{g}T_{ga}{}^{\g}+T_{\a\b}{}^{\la}T_{\la
 a}{}^{\g}+
 T_{a(\a|}{}^{\la}T_{|\b)\la}{}^{\g}-T_{a (\a|}{}^{g}T_{|\b)
 g}{}^{\g}\nn\\-\nabla_{(\a|}T_{|\b)a}{}^{\g} -\nabla_{a}T_{\a
 \b}{}^{\g}=0
\eea

Although not yet simplified this identity predicts the same term
that we found exist in $R^{(2)}{}_{\a~amn}$. However it includes a
great deal more information which we have included in another
letter.

\section{Conclusions}

We have found a consistent solution to the manifestly supersymmetric
equations of D=10, N=1 Supergravity, with string corrections to
second order in the string slope parameter. We have reaffirmed the
results of \cite{3} and \cite{4}, and we have solved the remaining
previously intractable curvature. We find a new and important
modification to the first order case as in equation (31). We gave
more careful consideration to the imposition of the constraint (9),
and we note that imposing this constraint will modify the solution.
However a solution can also be found by correspondingly modifying
the constraint (18). This solution allows for flexibility in finding
a suitable candidate for the supercurrent $A^{(2)}{}_{pqr}$.
Otherwise it is tied to the torsion $T^{(2)}{}_{\a b}{}^{\d}$

\section{Acknowledgement}
I would like to acknowledge S. Bellucci for introducing me to the
method of Bianchi identities and to recognize the founding work done
in this area by S. J. Gates Jr,.

\section{Appendix}

Here for convenience we list the torsions curvatures and H sector
results to second order. Other first order results listed in
\cite{1} also form part of the set. \bea H_{\a\b\g}=0 +
Order(\g^{3})\eea

\bea H_{\a\b d}=+\frac{i}{2}\s_{d \a\b}
+4i\g\s^{g}{}_{\a\b}H_{\g}{}^{ef}H_{d}{}^{ef}\nn\\
~\s_{\a\b}{}^{g}[{8i\g}H^{(0)}{}_{def}L^{(1)}{}_{g}{}^{ef}~ -~i\g
H^{(0)}{}_{def}A^{(1)}{}_{g}{}^{ef}]\nn\\~+~
\s^{pqref}{}_{\a\b}[\frac{i\g}{12}H^{(0)}{}_{def}A^{(1)}{}_{pqr}]+Order(\g^{3})
\eea

\bea H_{\a ab}=+2i\g[-\s_{[a|}{}_{\a\b}T_{ef}{}^{\b}G_{|b]}{}^{ef}-
2\s_{e}{}_{\a\b}T_{f[a|}{}^{\b}G_{|b]}{}^{ef}]\nn\\
~2\g[\nabla_{\a}(H^{(0)}_{[a|ef}H^{(0)}{}_{|b]}{}^{ef}~
-~\s_{ab\a}{}^{\f}\nabla_{\f}(H^{(0)}{}_{gef}H^{gef})]\nn\\
~+2i\g\s_{[a|\a\f}T_{ef}{}^{\f}\Pi^{(1)}{}_{|b]}{}^{ef}
-~2i\g\s_{ab\a}{}^{\la}\s_{g\la\f}T_{ef}{}^{\f}\Pi^{(1)gef}\nn\\
-~\frac{\g}{6}\s^{g}{}_{[a|\a}{}^{\f}\s_{|b]\la\f}T_{ef}{}^{\la}\Pi^{(1)}{}_{g}{}^{ef}
-~\frac{\g}{6}\s^{g}{}_{[a|\a}{}^{\f}\s_{g\la\f}T_{ef}{}^{\la}\Pi^{(1)}{}_{|b]}{}^{ef}\nn\\
~-~4\g R^{(1)}{}_{\a[a|}{}^{ef}H^{(0)}{}_{|b]ef}+T^{(2)}{}_{\a
ab}+Order(\g^{3})\eea

\bea T_{\a\b}{}^{g}=
i\s_{\a\b}{}^{g}-\frac{i\g}{6}\s^{pqref}{}_{\a\b}
H^{(0)}{}^{d}{}_{ef}A^{(1)}{}_{pqr}+Order(\g^{3})\eea

\bea
T_{\a\b}{}^{\g}=-[\d_{(\a|}{}^{\g}\d_{|\b)}{}^{\d}+\s^{g}{}_{\a\b}\s_{g}{}^{\g
\d}]\chi_{\d}~-~\frac{i\g}{12}\s^{pqref}{}_{\a\b}A^{(1)}{}_{pqr}T_{ef}{}^{\g}+Order(\g^{3})
\eea

\bea T_{\a g}{}^{\d} =
 -\frac{1}{48}\s_{g \a \f}\s^{pqr}{}^{\f\d}A^{(1)}_{pqr}
 +2\g~T^{(0)}{}^{ef}{}^{\d}
\Omega^{(1)}{}_{\a gef} +Order(\g^{3})\eea

\bea \s^{g}{}_{(\a\b|}T^{(2)}{}_{|\g)gd}
=4\g\s^{g}{}_{(\a\b|}\Omega_{|\g) gef}H^{(0)}{}_{d}{}^{ef}-
\frac{i\g}{6}\s^{g}{}_{(\a\b|}\s^{pqre}{}_{g|\g)\f}A^{(1)}{}_{pqr}T^{(0)}{}_{de}{}^{\f}
\eea
 Or symmetrized,

\bea T{}_{\g ab}=
+2\g[\Omega^{(1)}{}_{\g[a|ef}]H^{(0)}{}_{|b]}{}^{ef}
+\s_{ab~\g}{}^{\f}[\frac{\g}{3}\Omega^{(1)}{}_{\f
gef}H^{(0)}{}^{gef}]\nn\\
-\frac{\g}{6}\s_{[a|}{}^{g}{}_{\g}{}^{\f}\{\Omega^{(1)}{}_{\f
|b]ef}H^{(0)}{}_{g}{}^{ef}
+\Omega^{(1)}{}_{\f gef}H^{(0)}{}_{|b]}{}^{ef}\}\nn\\
-\frac{i\g}{12}A^{(1)}{}_{pqr}\s^{pqrg}{}_{[a|}{}_{\f\la}T^{(0)}{}_{|b]g}{}^{\la}\nn\\
-\frac{i\g}{72}\s_{ab~\g}{}^{\f}\s^{pqreg}{}_{\f\la}A^{(1)}{}_{pqr}T^{(0)}{}_{eg}{}^{\la}\nn\\
\frac{i\g}{144}A^{(1)}{}_{pqr}\s_{[a|}{}^{g}{}_{\g}{}^{\f}[\s^{pqre}{}_{|b]}{}_{\f\la}T^{(0)}{}_{eg}{}^{\la
}+\s^{pqre}{}_{g}{}_{\f\la}T^{(0)}{}_{e|b]}{}^{\la}]+Order(\g^{3})+...\nn\\
\eea

\bea R_{\a\b de} =
-2i\s^{g}{}_{\a\b}\Pi_{gde}{}^{(1)}+\frac{i}{24}\s^{pqref}{}_{\a\b}A_{pqr}{}^{(1)}\nn\\
-~\frac{i\g}{12}\s^{pqref}{}_{\a\b}A^{(1)}{}_{pqr}R_{ef
de}+Order(\g^{3})
 \eea
 Where

 \bea
 \Pi^{(1)}{}_{g}{}^{ef}=L^{(1)}{}_{g}{}^{ef}
-~\frac{1}{8}A^{(1)}{}_{g}{}^{ef} \eea

\bea
 R_{\a gde}= -i\s_{[d|}{}_{\a \f}T_{g |e]}{}^{f} +i\g\s_{[g|}{}_{\a~
 \f}T_{kl}{}^{\f}R^{kl}{}_{|de]}\nn\\
 + 2\g
 R^{(0)}{}_{abde}[\Omega^{(1)}{}_{\a
g}{}^{ab}]
+\nabla_{\a}\{\Pi^{(0)+(1)}{}_{gde}\}{}^{Order (2)}+Order(\g^{3})\nn\\
\eea

The spinor derivative of $L_{abc}$ is solved and available from a
Bianchi identity. We will list it in a later paper.

\end{document}